\documentclass{iopart}
\usepackage{graphicx}
\usepackage{amssymb}
\usepackage{color}
\usepackage{subfigure}
\expandafter\let\csname equation*\endcsname\relax                  
\expandafter\let\csname endequation*\endcsname\relax 
\usepackage{amsmath}

\usepackage[svgnames]{xcolor}
\usepackage{hyperref}
\usepackage{epsf}
\usepackage{psfrag}

\def \be{\begin{equation}}
\def \ee{\end{equation}} 

\begin{document}

\title{Gate-modulated thermopower of disordered nanowires: II. Variable-Range Hopping Regime}
\author{Riccardo Bosisio, Cosimo Gorini, Genevi\`eve Fleury and Jean-Louis Pichard\footnote{Corresponding author: jean-louis.pichard@cea.fr}}
\address{Service de Physique de l'\'Etat Condens\'e (CNRS URA 2464), IRAMIS/SPEC, CEA Saclay, 91191 Gif-sur-Yvette, France}

\begin{abstract}
We study the thermopower of a disordered nanowire in the field effect transistor configuration. After a first paper devoted to the elastic 
coherent regime (Bosisio R., Fleury G. and Pichard J.-L. 2014 \textit{New J. Phys.} \textbf{16} 035004), we consider here the inelastic activated regime taking place at higher temperatures. In the case where charge transport 
is thermally assisted by phonons (Mott Variable Range Hopping regime), we use the Miller-Abrahams random resistor network model as recently 
adapted by Jiang \emph{et al.} for thermoelectric transport. This approach previously used to study the bulk of the nanowire impurity band 
is extended for studying its edges. In this limit, we show that the typical thermopower is largely enhanced, attaining values larger that 
$10\, k_B/e \sim 1\, \mathrm{mV\,K}^{-1}$ and exhibiting a non-trivial behaviour as a function of the temperature. A percolation theory by Zvyagin 
extended to disordered nanowires allows us to account for the main observed edge behaviours of the thermopower.
\end{abstract}
  

\maketitle

\section{Introduction}
\label{sec_introduction}
The conversion of temperature to voltage differences or its inverse, enabling respectively waste heat recovery or cooling, 
is the purpose of a thermoelectric device. In linear response the device efficiency is controlled by its dimensionless figure 
of merit $ZT=S^2G T/\Xi$, with $T$ the temperature, $S$ the Seebeck coefficient or thermopower and $G, \Xi$ respectively 
the electrical and thermal conductances. As ZT increases, the efficiency moves closer to the Carnot limit. The stronger 
the particle-hole {\it asymmetry} in a system is, the higher $S$ will be. An ideal thermoelectric device should then exploit 
to the maximum such asymmetry, while at the same time ensuring a poor thermal and a good electrical conductance~\cite{Chen2003}. 
Whereas the former requirement is necessary to increase efficiency, the latter is needed for enough electric (cooling) power  
to be extracted from a heat engine (Peltier refrigerator). From this perspective, semiconductor nanowires appear as very promising 
central building blocks of flexible, efficient and environmentally friendly thermoelectric converters~\cite{Hicks1993,Curtin2012,
Blanc2013,Brovman2013,Stranz2013,Karg2013,Roddaro2013}. Whereas their thermoelectric properties can be easily tuned by 
gates~\cite{Brovman2013,Roddaro2013}, the phononic contribution to thermal transport $\Xi_{ph}$  is suppressed due to the 
reduced dimensionality~\cite{Curtin2012,Blanc2013}, and a good power output could be achieved by stacking them in 
parallel~\cite{Hochbaum2008,Curtin2012,Stranz2013}. Furthermore Si-based devices, already under intense investigation~\cite{Tilke2002,
Bourgeois2007,Boukai2008,Hochbaum2008,Galli2010,Heron2010,Curtin2012,He2012,Blanc2013,Stranz2013}, exploit an abundant and non-polluting 
resource.
  
Most existing works concentrate either on highly doped samples~\cite{Tilke2002,Boukai2008,Hochbaum2008,Curtin2012,Karg2013,Stranz2013} 
or on the thermal conductivity of undoped wires~\cite{Bourgeois2007,Heron2010,Hu2012,He2012,Blanc2013}.
On the other hand recent studies by Jiang \emph{et al.} \cite{Jiang2012,Jiang2013} have rekindled the interest for systems
in which electronic transport takes place via phonon-assisted hopping between localised states, of which disordered nanowires with 
low carrier density are a paradigmatic realization.
Whereas in our first paper~\cite{Bosisio2014} we focused on the low-temperature coherent regime, in this work we extend the approach reviewed in Refs.~\cite{Jiang2012,Jiang2013} in order to investigate band-edge transport in the (activated) hopping regime.
Two simple physical mechanisms have in this regime a synergy which is ideal for thermoelectric conversion~\cite{Zvyagin1973,Zvyagin1991}:
(i) a strongly broken particle-hole symmetry due to the Fermi level lying close to the band edge;
(ii) a wide energy window around the Fermi level made available for transport by the phonons. In other words, the phonons lend 
the carriers the energy necessary for them to hop through the system, but of the latter only one species, either electrons or holes, 
has available states and thus actually propagates.

\begin{figure}
\centering
  \includegraphics[keepaspectratio, width=0.75\columnwidth]{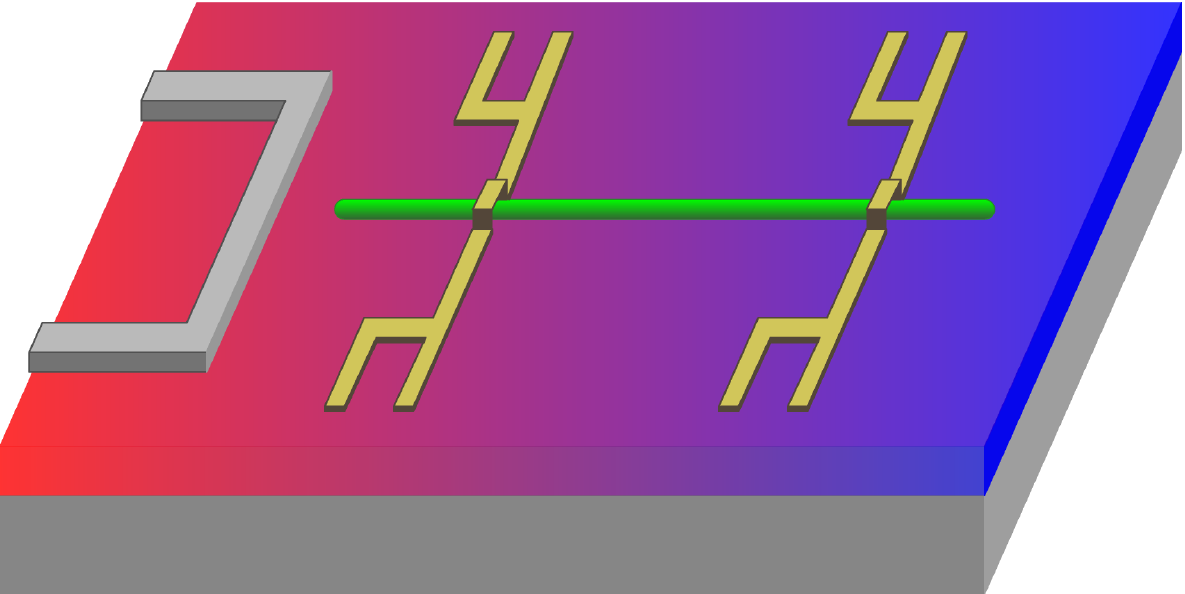}
  \caption{\label{Fig1} 
Nanowire in the field effect transistor (FET) device configuration: The nanowire (green) with two metal 
contacts (yellow) is deposited on an insulating substrate (blue). A heater (grey) makes the left side 
of the setup hotter (red) than its right side. A back gate (dark grey) is put below the substrate.}
\end{figure}

The general setup we have in mind is sketched in Fig.~\ref{Fig1}: A disordered semiconductor nanowire (green) connected to two 
metallic contacts (yellow) and deposited on an insulating substrate (blue). A heater (grey) and an applied bias voltage can induce a temperature and
an electrochemical potential difference between the two contacts. 
A back gate (dark grey), placed below the substrate, allows to shift the impurity band of the nanowire by means of a gate voltage. This way,
the transport of charges and heat can be studied when the Fermi potential of the setup probes either the bulk of the band or its edges.
Fig.~\ref{Fig1} depicts the more commonly used field effect transistor (FET) configuration~\cite{Brovman2013}. 
Another possibility would be to cover only the nanowire with a top gate (see e.g. Ref.~\cite{Poirier1999}). 
Putting a back gate is easier, but large gate voltages (a few hundreds volts) are necessary for shifting the impurity band, while few volts are 
sufficient if one uses a top gate. The nanowire itself could be (i) lightly doped, with electrons localised around distant impurity states, or (ii) 
highly doped but strongly depleted, or (iii) made of an amorphous semiconductor. A crucial feature of such wires is that their length $L$ 
should be much longer than the localisation length $\xi$ of their electron states, such that their electrical resistance becomes exponentially large when the temperature is lowered below a few Kelvin degrees. 
A crude modelling of such setup is sketched in Fig.~\ref{Fig2}: A purely 1D disordered chain with $L \gg \xi$, connected to two 
electron reservoirs and to a phonon bath (represented by the substrate), and coupled to a gate used to modulate its carrier density. Each site of the chain
corresponds to an electronic state localised by disorder or bound to impurity sites. Though such a model is strictly 1D, it should allow us to describe also quasi-1D wires ~\cite{Tilke1999,Dayen2009,Rodin2010} as long as their transverse sizes remain negligible compared to the typical hopping length (which we will define later).

In this work we are primarily interested in the thermopower $S$ of a single nanowire, as studied in many recent 
experiments~\cite{Brovman2013,Stranz2013,Karg2013,Roddaro2013}. We start in Sec.~\ref{section_transport} by introducing model and methods employed, moving on to discuss the nanowire 
electrical conductance in Sec.~\ref{section_conductance} and its thermopower in Sec.~\ref{section_thermopower}, before concluding in Sec.~\ref{section_ccl}. 
Various technical details, skimmed over in the main body for ease of reading, are gathered in the appendices.   

\section{Model and method}
\label{section_transport}
\begin{figure}
  \centering
  \includegraphics[keepaspectratio, width=0.9\columnwidth]{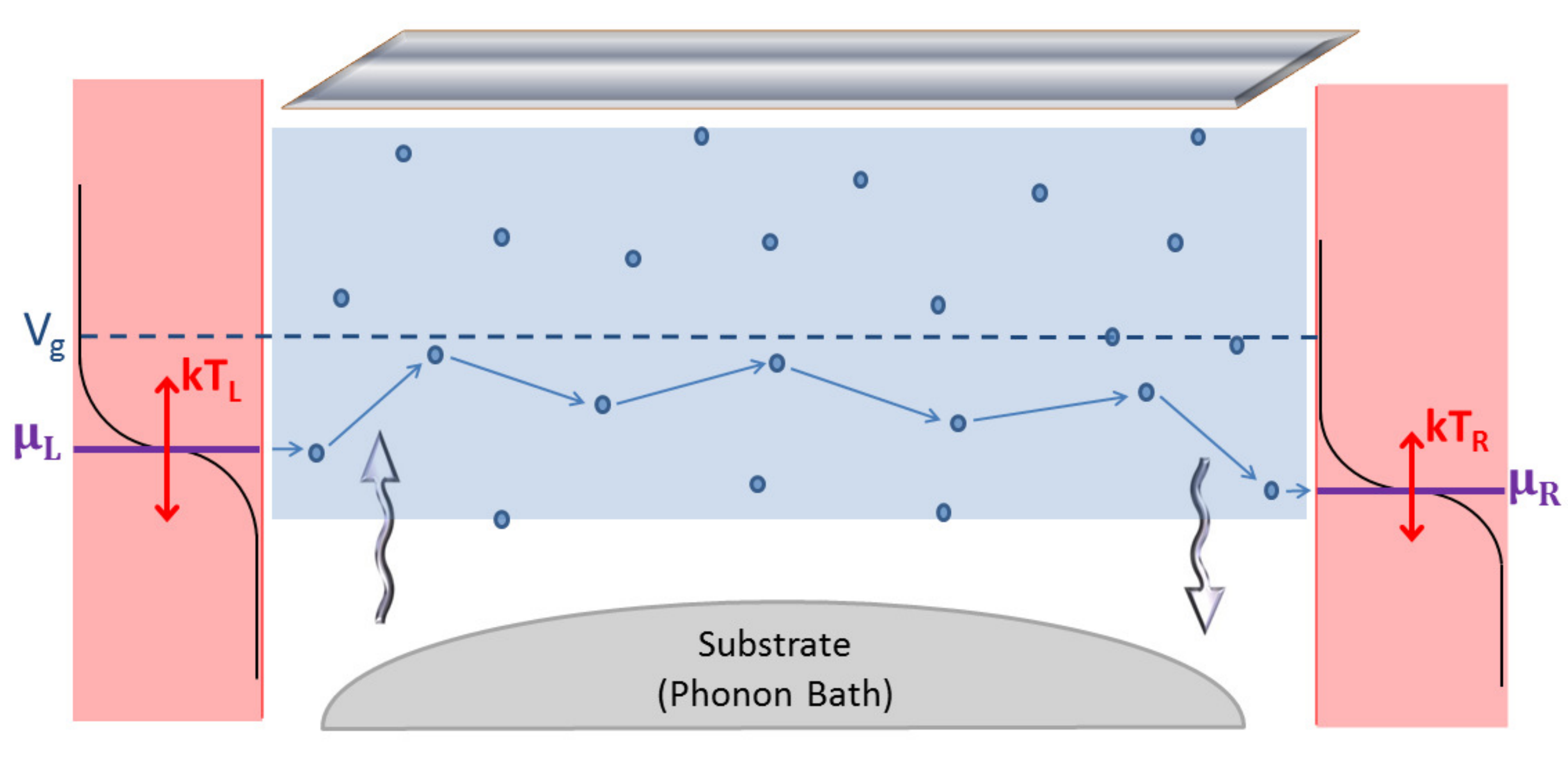}
  \caption{\label{Fig2} Variable Range Hopping (VRH) transport for a disordered nanowire in a FET configuration:
Two ohmic contacts are connected by a 1D disordered chain of length $L$ where the electron states are localised. 
The contacts are thermalised at temperatures $T_L(T_R)$ with electrochemical potentials $\mu_L(\mu_R)$ respectively. 
The electronic states (blue dots of coordinates ($x_i,E_i$)) are localised in regions of size $\xi_i \ll L$. 
Their centres $x_i$ are taken at random along the chain, with energies $E_i$  distributed inside an impurity band of width 
$2E_B$ (shaded light blue region). A top gate (in grey  at the top of the figure) allows to shift the impurity band. 
The gate potential $V_g$ sets the center of the band (dashed line). In linear response, the carriers are injected from the 
left (right) contacts inside the electronic states localised near the edges of the chain, in a window of energies of order 
$k_BT_L(k_BT_R)$ around $\mu_L(\mu_R)$. Inside the chain, the carrier propagation is thermally assisted by phonons (wavy arrows), which allow 
a carrier to do hops of variable range between localised states at different energies. The phonon bath at temperature 
$T_{ph}$ is represented by the substrate upon which the chain is deposited.}
\end{figure}

As sketched in Fig.~\ref{Fig2}, we consider a disordered nanowire of length $L$ in which all available electronic states are exponentially 
localised at positions $x_i$, with a localisation length $\xi_i\ll L$. We assume each state $i$ is either empty, 
or occupied by a single electron, but cannot be doubly occupied owing to a strong on-site Coulomb 
repulsion\cite{Ambegaokar1971}. The energy levels $E_i$ of the localised states are distributed within a band of width 
$2E_B$ and $\nu(E)$ denotes their density of states (DOS) per unit length at energy $E$. They can be shifted as a whole 
by an external gate voltage $V_g$. The nanowire is attached at its ends to two metallic contacts held at electrochemical 
potentials $\mu_L$ and $\mu_R$ and temperatures $T_L$ and $T_R$.  It is also coupled to a phonon bath at temperature $T_{ph}$ 
which provides the energy for electrons to hop between localised states. We focus on the situation in which the temperature $T$ 
is the same in all reservoirs ($T_L=T_R=T_{ph}\equiv T$) and consider linear response, assuming the difference in  electrochemical potentials between left 
and right leads to be small 
($\mu_L=\mu+\delta\mu\gtrsim\mu_R\equiv\mu$).\\

\subsection{Identification of the different transport mechanisms and of their temperature scales}
\label{subsec_Tscale}
Transport through the nanowire happens as follows.  Since there is a continuum of available states in the leads, 
we assume that charge carriers, let us say electrons, enter or leave the nanowire by elastic tunneling processes, without 
absorbing or emitting phonons\footnote{Phonon absorption and emission in the electrodes could be straightforwardly
taken into account.  However, it should not add any new physics and we neglect it.}. Inside the nanowire 
they have the possibility to hop either to localised states at higher energies by absorbing phonons, or to 
localised states at lower energies by emitting them.  Determining precisely the favoured electronic paths is a 
complicated task.  The proper way to tackle this issue is to map the hopping model to an equivalent 
random resistor network~\cite{Miller1960} and then to reduce it to a percolation problem~\cite{Ambegaokar1971}.
Such microscopic approaches are needed for giving precise quantitative predictions, but Mott's original 
argument~\cite{Mott1969,Mott1979} gives the main ideas: Assuming the localisation lengths and the density of states to be constant within a certain window of energies $\Delta$ to be explored ($\xi_i\approx \xi$, $\nu(E)\approx \nu$), the electron transfer from one localised state to 
another separated by a distance $x$ and an energy $\delta E \propto 1/(\nu x^D)$ ($D=1$ for us) results from 
a competition between the elastic tunneling probability $\propto \exp-(2x/\xi)$ to do a hop of length $x$ in 
space and the Boltzmann probability ($\propto \exp-(\delta E/k_BT)$) to do a hop of $\delta E$ in energy. 
Short hops are favoured by the former but are too energy-greedy for the latter, since localised states 
close in space are far in energy. This competition gives rise to an optimal electron hopping length, the 
Mott hopping length, which reads 
\be
\label{L_Mott}
L_M= \sqrt{\frac{\xi}{2 \nu k_BT}}
\ee 
in one dimension. 
$L_M$ is a decreasing function of the temperature, which allows us to define two characteristic temperature scales: 
the \textit{activation temperature}
\be
\label{T_Activation}
k_BT_x=\frac{\xi}{2\nu L^{2}}
\ee
at which $L_M\simeq L$ and the \textit{Mott temperature}
\be
\label{T_Mott}
k_BT_M=\frac{2}{\nu \xi}, 
\ee
at which $L_M\simeq \xi$. 
At low temperatures $T<T_x$, $L_M$ exceeds the system size and transport through the nanowire occurs via elastic 
coherent tunneling (see Ref.~\cite{Bosisio2014}). Above $T_x$, transport becomes inelastic, and remains coherent 
at scales smaller than $L_M$ only. The regime of intermediate temperature $T_x<T<T_M$ is known as the variable-range 
hopping (VRH) regime. As sketched in Fig.~\ref{Fig2}, electronic transport in this regime is achieved via several jumps of 
length $\approx L_M$ (with $\xi<L_M<L$). As it can be proven using a microscopic approach based on random resistor 
networks and percolation theory~\cite{Ambegaokar1971,Zvyagin1973,Zvyagin1991}, the VRH conductance can be simply 
expressed in terms either of $L_M$, $T_M$ or the hopping energy $\Delta$,
\be
\label{conductance_VRH}
G \propto \text{exp} \left\{-\frac{2L_M}{\xi}\right\}=\text{exp}\left\{-\sqrt{\frac{T_M}{T}}\right\}=\text{exp}\left\{-\frac{\Delta}{k_BT}\right\},  \\
\ee
where (it will be of prime importance later on)
\be
\Delta= k_B\sqrt{T_M T}
\ee
defines the width of the energy interval around $\mu$ inside which are located all states contributing to transport. 
Let us underline that if $T_x<T \ll T_M$, $\Delta$ becomes much larger than $k_B T$, the relevant energy interval for 
transport in the coherent regime ($T<T_x$). At large temperatures $T>T_M$, $L_M$ becomes of the order of or even smaller 
than the localisation length $\xi$, and one enters the nearest-neighbour hopping (NNH) regime where transport is 
simply activated between nearest neighbour localised states. Actually, in 1D, the crossover from VRH to 
simply activated transport is expected to take place at temperatures lower than $T_M$. The reason is the presence of highly resistive 
regions in energy-position space, where electrons cannot find empty states at distances $\sim \Delta, L_M$. These regions can be 
circumvented in 2D or 3D but not in 1D, where they behave as "breaks" in the percolating path: electrons are topologically constraint to 
cross them by thermal activation, making the temperature dependence of the overall resistance simply activated~\cite{Kurkijarvi1973,Raikh1989}.  
The critical temperature $T_a$ that marks the onset of this simply activated behaviour is given implicitly by the relation~\cite{Serota1986} 
\be
\label{T_a}
L= \frac{\xi}{2} \sqrt{\frac{T_M}{2T_a}}\, \text{exp}\left\{\frac{T_M}{2T_a}\right\}.
\ee
Below $T_a$, the probability of having such breaks in the nanowire can be neglected.\\
\begin{figure}[h!]
  \centering
  \includegraphics[keepaspectratio,width=0.75\columnwidth]{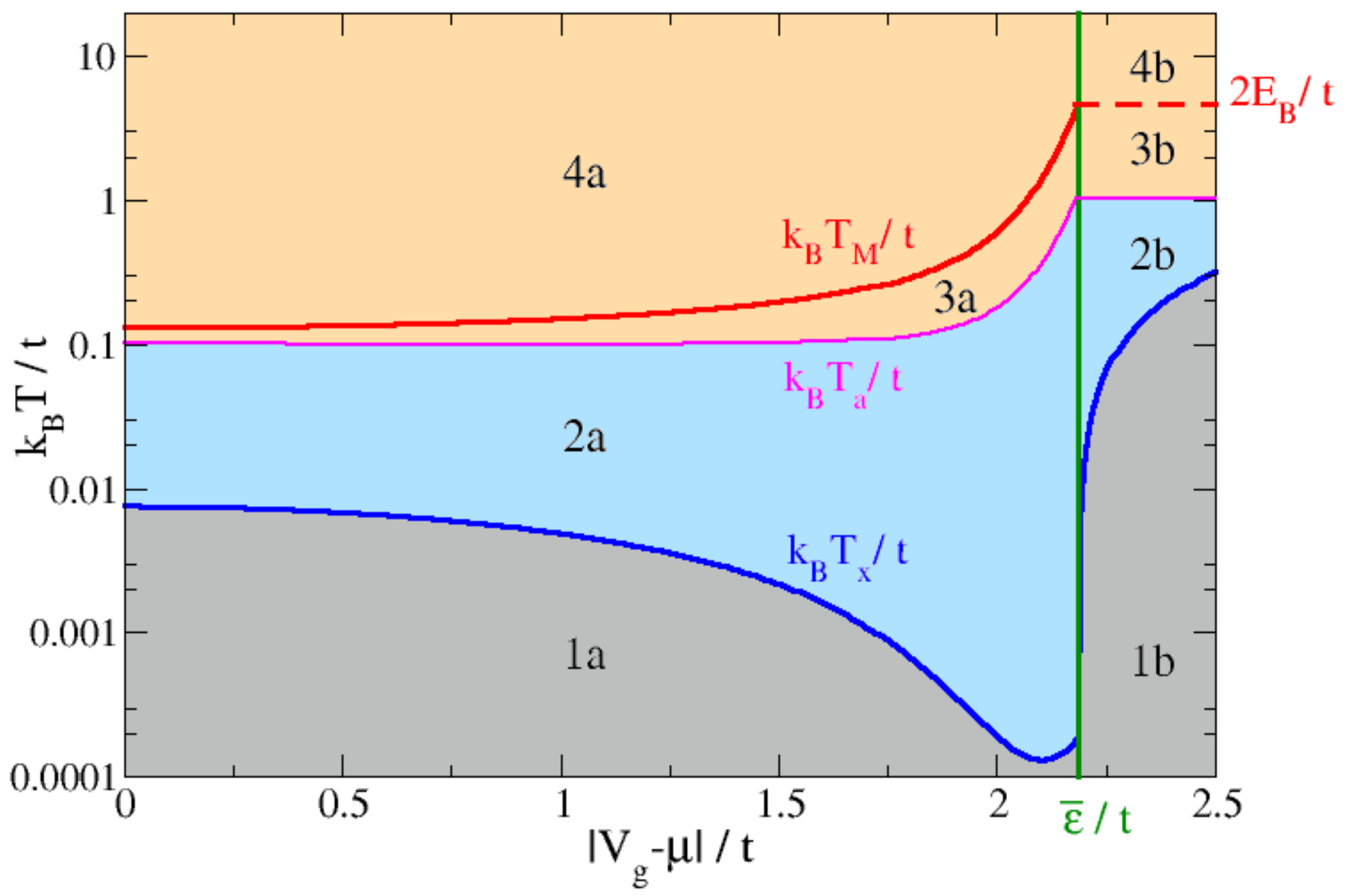}
  \caption{\label{Fig3}
  Gate dependence of the temperature scales separating the different regimes of electronic transport in a disordered nanowire: Elastic regime 
(grey), inelastic VRH regime (blue) and simply activated regime (red)). By varying the gate voltage $V_g$, one scans the impurity band, starting from its 
center (when $V_g-\mu=0$) towards its edges (approximately for $|V_g-\mu|=\bar{\epsilon}$) and ending up eventually outside the band (when $|V_g-\mu|\gtrsim 
\bar{\epsilon}$). The scales $T_x$, $T_a$ and $T_M$ defined in Sec.~\ref{subsec_Tscale} have been plotted for the Anderson model introduced in 
 Sec.~\ref{sec:Andmodel}, with $W=t$ and $L=200$.}
\end{figure}

\indent In Fig.~\ref{Fig3} the temperatures $T_x$, $T_M$ and $T_a$ are given as a function of the gate voltage $V_g$, taking for the disordered nanowire an Anderson model where the $L$ random site potentials are shifted by $V_g$\footnote{We assume the gate acting only along the nanowire, which corresponds to using a top gate. A FET configuration with a back gate should behave similarly, the field effect in the metallic contacts being negligible.}, the electrochemical potential $\mu$ being fixed in the reservoirs.

Still following Mott's approach, we consider $\xi_i \approx \xi$ and $\nu(E) \approx \nu$ (both evaluated at $\mu$), thus neglecting their variations within $\Delta$. The shape of the curves is a consequence of the explicite energy dependence of the localisation length $\xi$ and of the DOS $\nu$, which is detailed in Sec.~\ref{sec:Andmodel}.
Approaching an impurity band edge ($\pm E_B$), both $\xi$ and $\nu$ decrease rapidly, inducing a large increase of $T_M$ and $T_a$ that must be eventually cut-off when $T_M$ exceeds the bandwidth $2E_B$. Indeed, when $T\to T_M=2E_B$, $\Delta \to 2E_B$ and the range of states available for hopping transport reaches its limit.
More explicitly, we estimate this to happen at an energy scale $|\mu-V_g|\approx\bar{\epsilon}\approx 2.2t$ for the set of parameters considered in Fig.~\ref{Fig3}.

When Eqs.~\eqref{T_Activation},~\eqref{T_Mott}, and \eqref{T_a} cease to be valid, we will use a simplified model introduced by Zvyagin for estimating the temperatures $T_x$, $T_M$ and $T_a$. In this model, the DOS drops abruptly from a constant to $0$ at $\bar{\epsilon}$. This yields that, when $|\mu-V_g| \gtrsim \bar{\epsilon}$, $T_M$ and $T_a$ do not vary anymore and keep their values at $\bar{\epsilon}$, while the activation temperature $T_x$ gives the energy that electrons need in order to jump inside the band: $k_BT_x\approx |\mu-V_g|-\bar{\epsilon}$. We will show later that the 
edge behaviours numerically obtained using the Anderson model are well described by this simplified model.\\ 

\indent As a summary, let us now discuss the regimes of electronic transport corresponding to each region of the temperature diagram established 
in Fig.~\ref{Fig3}. Standard VRH regime takes place in region~(2a), at intermediate temperatures, 
when $\mu$ lies inside the impurity band. According to Mott law in 1D, the average logarithm of the resistance 
behaves there as $(T_M/T)^{1/2}$. In Sec.~\ref{section_conductance}, we will see how this statement has to be revisited 
in the vicinity of the band edges, and how to take into account the energy dependency of $\xi$. At higher temperatures, transport is simply activated (the temperature dependence of the logarithm of the resistance $\propto T^{-1}$).  
This is due either to the presence of a very resistive link in the best conducting path that dominates the resistance (region~(3a)), or simply 
to the fact that the thermal energy $k_B T$ is so high that transport occurs via hops between nearest neighbour states, no matter how far in energy 
they are (region~(4a)).  On the contrary at lower temperatures, in region~(1a), $L \le L_M$ and transport ceases to be thermally activated to become 
elastic and coherent through the whole nanowire. The thermopower in this regime has been studied in Ref.~\cite{Bosisio2014}. If now $\mu$ lies outside the impurity band, electrons need to 
absorb energy in order to enter the band.  In region~(1b), $k_B T$ is too small for that (the only way for electrons to cross the nanowire 
is then to tunnel directly from one reservoir to the other, which results in a exponentially vanishing conductance).  
At higher temperatures, in regions (2b), (3b) and (4b), electrons can be thermally activated.  
Once they have entered the nanowire, they hop from site to site according to the mechanism prevailing 
in regions (2a), (3a) and (4a) respectively.\\

\subsection{Formulation in terms of a random resistor network}
\indent We follow the approach used in Refs.~\cite{Jiang2012,Jiang2013} for studying 
thermoelectric transport in the hopping regime. It consists in solving the Miller-Abrahams 
resistor network~\cite{Miller1960} which was first introduced for describing charge transport 
in weakly doped crystalline semiconductors and later on extended to non crystalline  Anderson insulators. 
The nodes are given by the localised states. Each pair of nodes $i, j$ is connected by an effective resistor, 
which depends on the transition rates $\Gamma_{ij}, \Gamma_{ji}$ induced by local electron-phonon interactions. 
In addition, one needs to connect this network to the leads, if one wants to calculate the charge and heat 
currents flowing through it. Usually (and actually, we did not find a reference where this is 
not the case) one assumes for calculating these transition rates that $\xi_i=\xi(E_i)\equiv \xi$ (evaluated at $\mu$) for the localisation lengths of the different states, which can be done if 
the variations of the $\xi_i$ are negligible within $\Delta$. Here we need to go beyond such an 
approximation, since we are interested in band edge transport, where those variations cannot be neglected. 
The procedure is summarized below.\\
\indent Let us consider a pair of localized states $i$ and $j$ of energies $E_i$ and $E_j$.  
Assuming no correlations between their occupation numbers, the (time-averaged) transition rate 
from state $i$ to state $j$ is given by the Fermi golden rule as~\cite{Jiang2013}
\be
\label{eq:gamma_ij}
\Gamma_{ij}=\gamma_{ij}\,f_i\,(1-f_j)\,\left[N_{ij}+\theta(E_i-E_j)\right]\,,
\ee
where $f_i$ is the average occupation number of state $i$ and $N_{ij}=[\text{exp}\{|E_j-E_i|/k_BT\}-1]^{-1}$ 
is the phonon Bose distribution at energy $|E_j-E_i|$.  The presence of the Heaviside function accounts for the 
difference between phonon absorption and emission~\cite{Ambegaokar1971}. $\gamma_{ij}$ is the hopping probability 
$i \to j$ due to the absorption/emission of one phonon when $i$ is occupied and $j$ is empty. Assuming that the energy dependence of $\xi$ can be neglected, in the limit $x_{ij}\gg\xi$ one obtains
\be
\label{eq:gamma_ij_2}
\gamma_{ij}\simeq \gamma_{ep}\,\text{exp}(-2x_{ij}/\xi)\,.
\ee
Here $x_{ij}=|x_i-x_j|$ is the distance between the states, whereas  $\gamma_{ep}$, containing the electron-phonon 
matrix element, depends on the electron-phonon coupling strength and the phonon density of states. Since it is weakly 
dependent on $E_i$, $E_j$ and $x_{ij}$ compared to the exponential factors, it is assumed to be constant. Under the widely 
used approximation~\cite{Ambegaokar1971,Pollack1972,Shklovskii1984,Zvyagin1973} $|E_{ij}| \gg k_BT$, Eq.~\eqref{eq:gamma_ij} 
reduces to:
\be 
\label{eq:gamma_ij_3}
\Gamma_{ij}\simeq \gamma_{ep}\,e^{-2x_{ij}/\xi}\,e^{-(|E_i-\mu|+|E_j-\mu|+|E_i-E_j|)/2k_BT}\,.
\ee
Hereafter, we will go beyond these standard approximations by considering the exact expression~\eqref{eq:gamma_ij} for 
$\Gamma_{ij}$, and by taking 
\be
\gamma_{ij} = \gamma_{ep}\left(\frac{1}{\xi_i}-\frac{1}{\xi_j}\right)^{-2} 
\left(\frac{\exp\{-2r_{ij}/\xi_j\}}{\xi_i^2}+\frac{\exp\{-2r_{ij}/\xi_i\}}{\xi_j^2} 
-\frac{2\exp\{-r_{ij}(1/\xi_i+1/\xi_j)\}}{\xi_i \xi_j}\right),  
\label{eq:gamma_ij_app_4}
\ee
for $\gamma_{ij}$. Eq.~\eqref{eq:gamma_ij_app_4} takes into account the energy dependence of $\xi(E)$ and is derived in~\ref{app_resnet}. 
\\
\indent The tunneling transition rates between each state $i$ and the leads $\alpha$ ($\alpha=L$ or $R$) are written in a similar way as 
\be
\label{eq:gamma_il}
\Gamma_{i\alpha}=\gamma_{i\alpha}\,f_i\,\left[1-f_{\alpha}(E_i)\right]
\ee
where
\be
\gamma_{i\alpha}\simeq\gamma_{e}\,\text{exp}(-2x_{i\alpha}/\xi_i)\,.
\ee
In the above equations $f_{\alpha}(E)=[\text{exp}\{(E-\mu_\alpha)/k_BT\}+1]^{-1}$ is lead $\alpha$'s
Fermi-Dirac distribution, $x_{i\alpha}$ denotes the distance of state $i$ from lead $\alpha$ 
and $\gamma_{e}$ is a rate quantifying the coupling between the localized states and the leads 
(taken constant for the same reason as $\gamma_{ep}$).\\
\indent Then, the net electric currents flowing between each pair of localized states and 
between states and leads are obtained by
\begin{subequations}
\label{eq:currents}
\begin{align}
I_{ij} &= e\,(\Gamma_{ij}-\Gamma_{ji})\label{eq:currents_1}\\
I_{i\alpha} &= e\,(\Gamma_{i\alpha}-\Gamma_{\alpha i})\,\qquad\,\alpha=L,R\label{eq:currents_2}
\end{align}
\end{subequations}
$e<0$ being the electron charge. The linear response solution of this random resistor network problem is reviewed in Ref.~\cite{Jiang2013}.  Details of the calculation of the charge currents
and heat currents are summarized in \ref{app_resnet} for the Peltier configuration we consider, 
where the temperature is $T$ everywhere and the reference (equilibrium) electrochemical potential is that of the right reservoir ($\mu\equiv \mu_R$). 
In this case the electrical conductance $G$, Peltier coefficient $\Pi$ and thermopower $S$ are determined within the Onsager formalism by the charge ($I^e_L$) and heat ($I^Q_L$) currents exchanged with the left reservoir:
\begin{subequations}
\label{eq:G_S_tot}
\begin{align}
    & G = \frac{I^e_L}{\delta\mu/e}  \label{eq:G_S_1},\\
    & \Pi = \frac{I^Q_L}{I^e_L} \label{eq:G_S_3},\\
    & S = \frac{\Pi}{T}=\frac{1}{T}\,\frac{I^Q_L}{I^e_L}. \label{eq:G_S_2}
\end{align}
\end{subequations}
In the last equation, the Kelvin-Onsager symmetry relation~\cite{Callen1985} $\Pi=ST$ has been used for deducing the thermopower. Notice that a different choice of reference could be adopted (see for instance Ref. \cite{Jiang2013}) without affecting the transport coefficients $G$, $\Pi$ and $S$.
 
\subsection{Anderson model for the localised states}
\label{sec:Andmodel}
The set of energies $E_i$ and localisation lengths $\xi_i$ are required as input parameters of the random resistor 
network problem.  To generate them we use the Anderson model.  The disordered nanowire is modeled as a 1D lattice of length $L$ with a lattice spacing $a$ set equal to one, described by a $L\times L$ tight-binding Hamiltonian:
\be
\label{eq:modelAnderson1D}
\mathcal{H}=-t\sum_{i=1}^{L-1}\left(c_i^{\dagger}c_{i+1}+\text{h.c.}\right)+\sum_{i=1}^{L}(\epsilon_i+V_g) c_i^{\dagger}c_i\,,
\ee
where $c^{\dagger}_i$ and $c_i$ are the electron creation and annihilation operators on site $i$ and $t$ 
is the hopping energy. In the following all energies will be expressed in units of $t$.  
The disorder potentials $\epsilon_i$ are (uncorrelated) random numbers uniformly distributed 
in the interval $[-W/2,W/2]$.  The constant potential $V_g$ is added to take into account 
the presence of an external top gate, allowing to shift the whole nanowire impurity band.\\
\indent By diagonalizing the Hamiltonian~\eqref{eq:modelAnderson1D}, we find the energies $E_i$ of 
the localised states. They are distributed with the DOS $\nu(E)$ in the interval $[V_g-E_B,V_g+E_B]$, 
$\pm E_B$ being the band edges of the model at $V_g=0$. In the limit $L\to\infty$, $E_B=2t+W/2$.  
To generate the localisation lengths $\xi_i$, we neglect sample-to-sample fluctuations and 
assume that $\xi_i$ is given by the typical localisation length $\xi(E_i)$ at energy $E_i$, 
characterizing the exponential decay of the average logarithm of the elastic conductance 
($\ln G\sim -2L/\xi$). The DOS $\nu(E)$ and localisation length $\xi(E)$ are shown in Fig.~\ref{Fig4}; 
their energy dependence is analytically known in the large size and small disorder limits, 
both in the bulk of the band and close to the edges (see Refs.~{\cite{Derrida1984,Bosisio2014}). Obviously, 
if $\mu$ lies close to the band edges and/or if the available energy window $\Delta$ around $\mu$ is not small compared 
to $t$, the energy dependency of $\nu(E)$ and $\xi(E)$ cannot be neglected.  This explains why we need to go beyond 
the approximation of constant DOS and localisation length when scanning the impurity band with the gate voltage.\\
\indent Solving the Anderson model gives us the full set of localised states: their energy levels $E_i$,
their localisation lengths $\xi_i=\xi(E_i)$ and their positions along the disordered chain. However, to speed up the procedure 
of building a basis of localised states, we simply assign the levels $E_i$ to random positions $x_i$ between $0$ and $L$ along 
the chain (with a uniform distribution).  This approximation is conventional in numerical simulations of VRH transport 
(see~\cite{Lee1984,Serota1986,Jiang2013} among others)~\footnote{By doing this we lose a feature of Anderson model, namely 
that states which are close in energy are distant in space, and as a consequence our model may overestimate the hopping 
between certain pairs of states.  However this should not play an important role if $L$ is sufficiently 
large, $L\gg\xi(\mu)$.  In this case states which are accidentally taken close both in space and in energy should be not only rare 
but, more importantly, can merely be seen -- regarding percolation -- as one small localised cluster, i.e. as a {\textit single} 
new effective localised state.  The reason is that the optimal percolation path is eventually determined by the most resistive 
links.  Thus, we can always reformulate the problem in order to end up in a situation in which neighbouring states are far away in 
energy.}.\\
\indent Hereafter, we will study disordered chains with a disorder strength $W=t$, which is sufficiently small for using 
weak disorder expansions~\cite{Bosisio2014} and sufficiently large for ensuring $L \gg \xi_i$ at relatively small 
sizes. For $V_g=0$ and $L\approx 1000$, the spectrum edges of the disordered nanowire are found at $E_B \approx 2.35\,t$, which 
is smaller than $2.5\,t$, the value characterizing the limit $L \to \infty$. Such finite size effects are a consequence of the 
infinitely small tails of the asymptotic DOS $\nu(E)$ shown in Fig.~\ref{Fig4}: States of energy close to $2.5\,t$ 
can only exist in infinitely long chains.

\begin{figure}
  \centering
  \includegraphics[keepaspectratio, width=0.75\columnwidth]{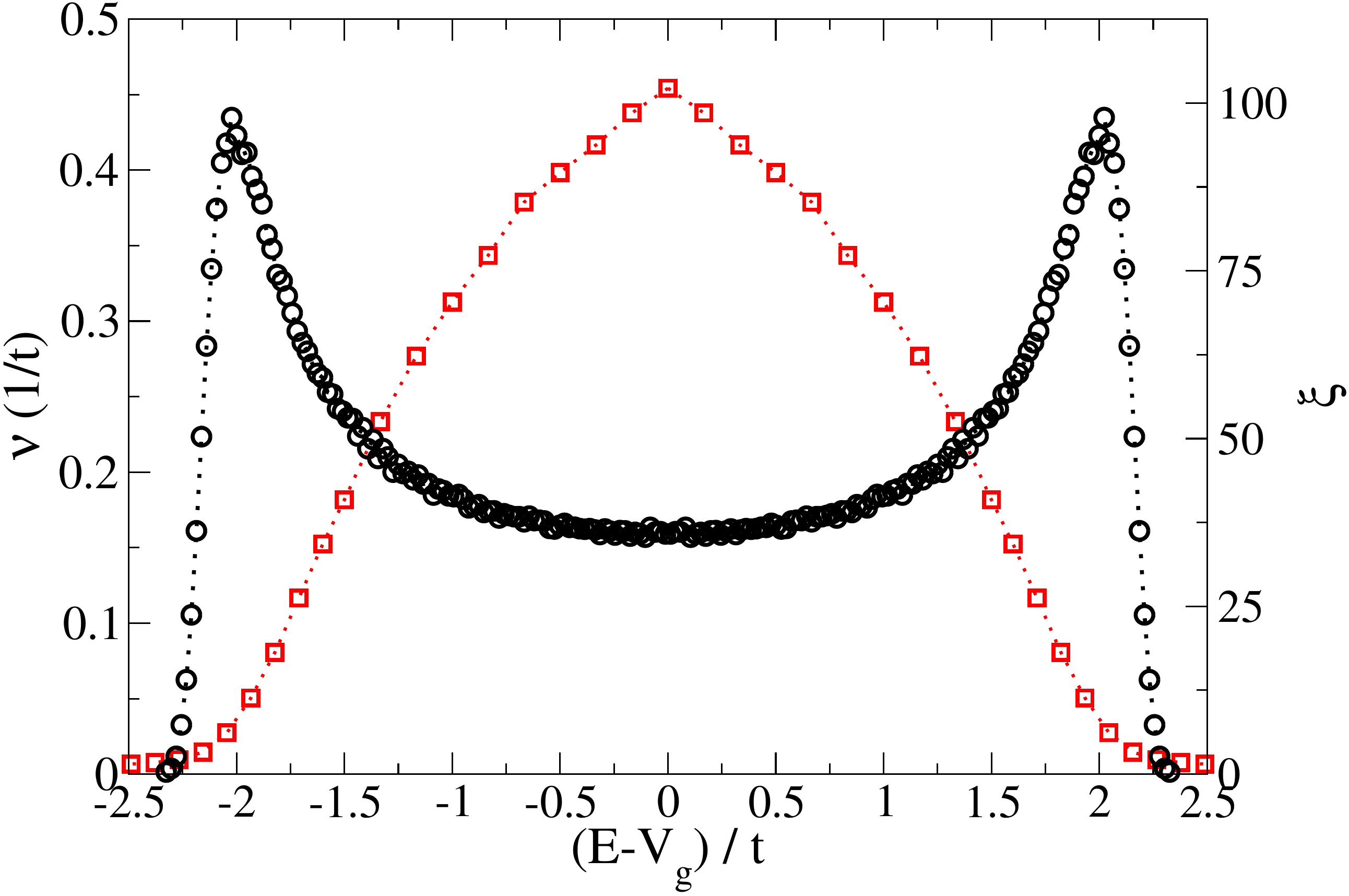}
  \caption{\label{Fig4} 
Density of states per site $\nu$ ({\large$\circ$}) and localisation length $\xi$ ({\scriptsize{\color{red}$\square$}}), 
as a function of energy $E$ for the 1D Anderson model \eqref{eq:modelAnderson1D} with disorder amplitude $W/t = 1$. 
The points correspond to numerical data (obtained in the large length limit with $L = 1600$). Analytical expressions describing 
$\nu((E-V_g)/t)$ and $\xi((E-V_g)/t)$ are given in Ref. \cite{Bosisio2014}.}
\end{figure}


\section{Electrical Conductance}
\label{section_conductance}

\subsection{Background}
The electrical conductance of one-dimensional conductors in the VRH regime has been much studied 
in the literature, both experimentally~\cite{Kwasnick1984,Webb1985,Ladieu1993exp,Hasko1993,Han2010} 
and theoretically~\cite{Kurkijarvi1973,Lee1984,Serota1986,Raikh1989,Ladieu1993th,Rodin2011}.  
In particular, the validity of Mott law for the typical conductance in 1D
\be
\label{eq:log_G_Mott}
\ln G(T) \sim -\alpha\sqrt{\frac{T_M}{T}},
\ee
with $\alpha\approx 1$, was a subject of controversy for a long time since strictly speaking, 
Mott's argument leading to Eq.~\eqref{eq:log_G_Mott} does not hold in 1D.  
It was shown that due to the presence of "breaks", the prefactor $\alpha$ is actually 
also a function of the temperature and system length~\cite{Serota1986,Raikh1989}.  
Nevertheless, the $T$- and $L$-dependency of $\alpha$ turns out to be so weak that at low temperatures 
$\alpha$ is almost constant and Mott law is recovered.  Taking the proper $\alpha(L,T)$ 
into account allows an analytical description of the crossover from Mott law to the activated behaviour,
$\ln G(T)\sim T^{-1}$, above $T_a$ (see Sec.~\ref{subsec_Tscale}) but the refinement thus introduced 
is too small to be clearly evidenced by numerical simulations and even less by experimental measurements.\\
\indent Another limitation of Mott's standard argument and of subsequent, more elaborate 
percolation-based ones is the initial assumption of a constant DOS and a constant localisation length 
around $\mu$.  As long as $\nu(E)$ is slowly varying in the energy window $|E-\mu|<\Delta$ 
(still keeping $\xi$ constant), Eq.~\eqref{eq:log_G_Mott} is expected to hold, but 
it lacks justification in the case of strongly varying DOS. In particular, Eq.~\eqref{eq:log_G_Mott} has to be 
revised when transport through the system occurs at energies around the impurity band edges.  
This question was tackled by Zvyagin in Refs.~\cite{Zvyagin1973,Zvyagin1991}, 
by approximating the DOS by a step-like function. If one considers the lower band edge, the approximated DOS 
reads
\be
\nu(E)\simeq \nu_0\, \theta(E-\epsilon_c),
\label{eq:dos_theta}
\ee
where $\epsilon_c$ plays the role of an effective band edge. Though \textit{three-dimensional} 
systems were considered in Refs.~\cite{Zvyagin1973,Zvyagin1991}, a similar approach can be extended 
to our 1D model setting $\epsilon_c=V_g-\bar{\epsilon}$, where $\bar{\epsilon}$ is the 1D effective edge 
introduced in Sec.~\ref{subsec_Tscale} for $V_g=0$. The idea is that when $\mu$ lies outside the impurity band, 
electrons need an activation energy $\epsilon_c-\mu$ in order to ``jump'' inside it to find available states.  
This entails an extra term in Eq.~\eqref{eq:log_G_Mott}, which in 1D becomes
\be
\label{eq:log_G_hybrid}
\ln G (T)\sim -\frac{E_A}{k_BT}-\bar{\alpha}\sqrt{\frac{T_M}{T}},
\ee
with $E_A\sim \epsilon_c-\mu$ and $\bar{\alpha}$ differing from $\alpha$ by some numerical factors~\cite{Grant1974,Zvyagin1991}.

\subsection{Numerical results}
\indent We have investigated numerically how the typical conductance of a disordered nanowire depends on 
the temperature when the applied gate voltage is varied. For the model described in Sec.~\ref{sec:Andmodel}, 
we have solved the random resistor network problem and calculated the conductance 
$G$ via Eq.\eqref{eq:G_S_1}. This procedure has been iterated over many random configurations 
of the energy levels $E_i$ in order to extrapolate the \emph{typical} logarithm of the conductance $[\ln G]_0$, 
defined as the median of the resulting distribution $P(\ln G)$\footnote{More details concerning 
the distributions of the logarithm of the conductance for 1D systems in VRH regime can be found 
in Refs.~\cite{Serota1986,Rodin2009}}. \\
\indent In Fig.~\ref{Fig5}, $[\ln G]_0(T)$ is plotted for two values of $V_g$, corresponding 
to the bulk and the lower edge of the band.  In both cases we show that low temperature 
data exhibit Mott law $T^{-1/2}$ behaviour (red dashed curve), while at higher temperatures 
they are well fitted by an activated $T^{-1}$ law (green dashed curve).  Eq.~\eqref{eq:log_G_hybrid} with 
adjusted values for $E_A$ and $\bar{\alpha}$ describes the crossover between the two regimes (full blue line). 
More precisely, when $\mu$ lies inside the band (Fig.~\ref{Fig5}(a)), 
the validity range of Mott law ($k_BT/t\lesssim 0.05$) is consistent with the required 
hypothesis of weakly varying DOS.  Indeed, below such temperatures, the energy window $\Delta=k_B\sqrt{T_MT}$ 
of accessible states around $\mu$ is so small ($\Delta\lesssim 0.2$ using for $T_M$ the value 
given in Fig.~\ref{Fig3}) that the DOS can be considered as weakly energy dependent 
($\Delta\left.\partial_E\ln \nu(E)\right|_{\mu}\approx 0.3<1$). This justifies the validity 
of Eq.~\eqref{eq:log_G_Mott} in such a regime.  
Note that the onset of activated behaviour at $k_BT\approx0.05\, t$ is also in rough agreement 
with the predicted value of $k_BT_a\approx 0.1\,t$ in Fig.~\ref{Fig3}.  
On the other hand, when $\mu$ lies in a region where the DOS is exponentially 
small (Fig.~\ref{Fig5}(b)), there is no more reason to use Mott law to describe 
our data, even if it appears to be well fitted by Eq.~\eqref{eq:log_G_Mott} at low temperatures.  
The point is that other power law formula, $[\ln G]_0\sim T^\beta$, could be used 
to fit our data in this narrow temperature range.  Thus, one cannot use the apparent suitability 
of Eq.~\eqref{eq:log_G_Mott} to support the validity of Mott law in this regime.  
Outside the band the correct framework for analysis is provided by Eq.~\eqref{eq:log_G_hybrid}.  
The activated contribution to the conductance is always present, which explains why 
in Fig.~\ref{Fig5}(b) the $T^{-1}$ fit starts to be accurate much below the temperature
$k_BT_a/t\approx 0.95$. Finally, at very high temperatures (typically larger than $t$), 
the typical conductance is found in both cases to decrease with temperature.  
This is due to the fact that in the limit $T\to\infty$, the factors $f_i(1-f_j)$ and $f_j(1-f_i)$ on one hand, 
and $f_i(1-f_\alpha)$, $f_\alpha(1-f_i)$ on the other, converge to the same value.  
Hence, the opposite rates $\Gamma_{ij}$, $\Gamma_{ji}$ and $\Gamma_{iL}$, $\Gamma_{Li}$ tend to level out, 
which results in a vanishing net current and a divergent resistance. An expansion of the Fermi functions to the 
next order in inverse temperature yields $I_{ij},\,I_{i\alpha}\sim T^{-1}$, which explains the linear decay at 
high $T$ of $[\ln G]_0$ versus $\ln T$ in Fig.~\ref{Fig5} (not marked).
\begin{figure}
  \centering
  \includegraphics[keepaspectratio, width=\columnwidth]{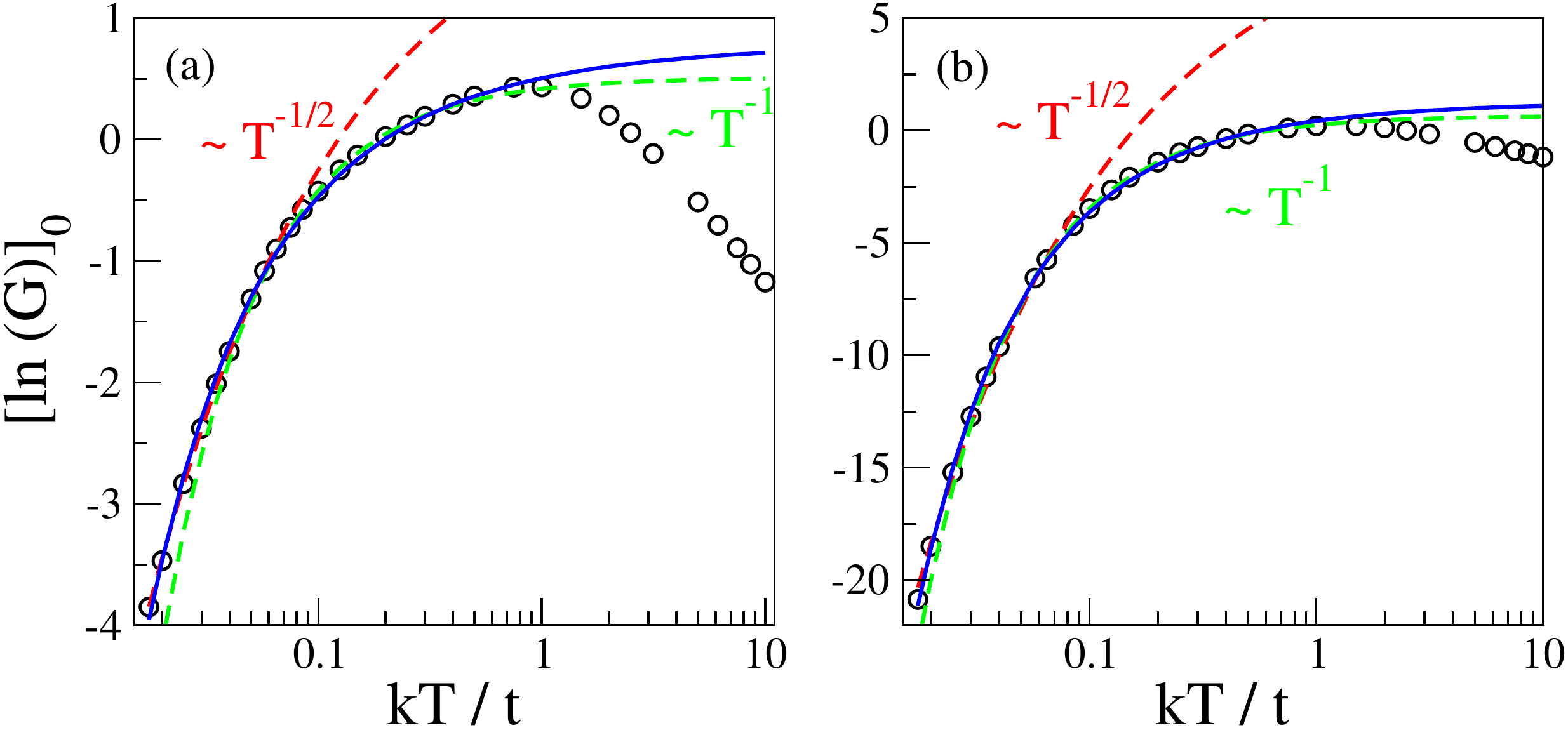}
  \caption{\label{Fig5} 
 Typical value of the logarithm of the conductance $[\ln G]_0$ as a function of $T$ for $\mu=0$ and two values of the gate 
voltage: (a) $V_g=1.9\,t$ inside the band and (b) $V_g=2.3\,t$ at the edge of the band. In both cases, at low temperatures, 
numerical data (points) are well fitted by a $T^{-1/2}$ fit (red dashed lines), evolving to a $T^{-1}$ behaviour as $T$ increases 
(green dashed lines). Full blue lines correspond to Eq.\eqref{eq:log_G_hybrid}, which describes the crossover between the two regimes. 
Parameters: $L=200$, $W=t$ and $\gamma_e=\gamma_{ep}=t$.}
\end{figure}

\section{Thermopower}
\label{section_thermopower}

\subsection{Background}
The thermopower is a measure of the average energy $\langle E-\mu \rangle$ transferred by charge 
carriers from the left lead to the right one.  In the low temperature coherent regime~\cite{Bosisio2014}, 
transport takes place near the Fermi energy. Hence, in linear response with respect to the bias voltage
between the two leads, the thermopower depends on the electron-hole asymmetry at $\mu$.  
On the contrary, in the VRH regime, all states in the energy window $|E-\mu|<\Delta$ contribute.  
Since $\Delta\gg k_BT$ when $T \ll T_M$, the thermopower benefits from the contribution of states far below 
and above $\mu$, despite being in linear response. When the gate voltage is adjusted in order to probe the 
impurity band edges, the electron contribution dominates over the hole one (or vice-versa), yielding 
an enhanced thermopower.\\
\indent To study the thermopower in the VRH regime~\footnote{
We stress that the usual Mott formula for the thermopower, $S=(\pi^2k_B^2T/(3e))\left.\partial_{E}\ln \sigma\right|_\mu$ 
($\sigma$ being the electrical conductivity), does not apply in the VRH regime, as pointed out by Mott himself in~\cite{Mott1979}. 
Indeed, this formula has been derived by averaging $\langle E-\mu \rangle$ within the standard Boltzmann formalism, not suitable 
in the VRH regime where $\Delta\gg k_BT$.}, we use the approach introduced by Zvyagin in~\cite{Zvyagin1973,Zvyagin1991}.  
The starting point is the percolation theory of hopping transport, according to which transport through the system 
is achieved via percolation in energy-position space.  The average $\langle E-\mu \rangle$ is calculated 
by averaging the energy over the sites composing the percolation cluster, and the thermopower is given by
\be\label{eq:S_zvyagin_1}
S=\frac{\langle E-\mu \rangle}{e\,T}=\frac{1}{e\,T}\,\frac{\int\!dE \,(E-\mu)\,\nu(E)\,p(E)}{\int\!dE \,\nu(E)\,p(E)},
\ee
where $p(E)$ is the probability that a state of energy $E$ belongs to the percolation cluster.  
The latter quantity is supposed to be proportional to the average number of bonds $N_b(E)$, given by
\be
\label{eq:N_bonds}
N_b(E)=\int\!dx \int\!dE'\,\nu(E')\,\theta\left(\sqrt{\frac{T_M}{T}}-\frac{2x}{\xi}-\frac{|E-\mu|+|E'-\mu|+|E-E'|}{2k_BT}\right),
\ee 
under the assumptions leading to Eq.~(\eqref{eq:gamma_ij_3}) 
($\mu$ inside the band, low temperature and energy independent localisation length $\xi(E)=\xi(\mu)$)~\cite{Ambegaokar1971,Zvyagin1991}. 
The Heaviside function $\theta$ accounts for the existence of a percolating path, and restricts the energy range of integration to the 
window $[\mu-\Delta,\mu+\Delta]$. After integrating over the single spatial variable $x$ (in 1D), one gets
\begin{align}
\label{eq:p_zvyagin}
p(E) \propto &\,\, \theta\left( \Delta - |E-\mu| \right)\times \cr 
& \int_{\mu-\Delta}^{\mu+\Delta}\!\!dE'\,\nu(E')\left(1-\frac{|E-\mu|+|E'-\mu|+|E-E'|}{2\Delta}\right)\theta\left(\Delta - |E-E'|\right).
\end{align}
Note that if $\mu$ lies outside the impurity band, electrons need to jump inside the latter by thermal 
activation before accessing the percolation cluster. In that case, Eqs.~\eqref{eq:N_bonds} 
and~\eqref{eq:p_zvyagin} have to be modified accordingly, by replacing $\mu$ by the energy $\epsilon_c$ 
of the closest band edge and by changing the energy range of integration to 
$[\epsilon_c,\epsilon_c+\Delta]$ (lower band edge) or $[\epsilon_c-\Delta,\epsilon_c]$ (upper band edge).\\
\indent Eqs.~\eqref{eq:S_zvyagin_1} and~\eqref{eq:p_zvyagin} enable us to calculate the thermopower 
once the DOS $\nu(E)$ is known. Following Zvyagin's works~\cite{Zvyagin1973,Zvyagin1991}, 
we discuss below a few extreme cases where the DOS takes a simple form. Contrary to those works focused 
on three-dimensional bulk materials, we derive expressions for the thermopower of nanowires in the 1D case.  
Despite the simplicity of our approach, we will see in the next subsection that it enables us to qualitatively 
capture the typical behaviour of the thermopower and the role of the gate (see Sec.~\ref{section_transport}).\\ 
\indent Let us first consider the case where \textit{(i)} the DOS can be approximated 
by its first order expansion $\nu(E)\approx \nu(\mu)+(E-\mu)\left.\partial_E\ln \nu(E)\right|_{\mu}$ in the 
interval $[\mu-\Delta,\mu+\Delta]$, and \textit{(ii}) $\nu$ is expected to vary slowly at the scale of $\Delta$, 
\textit{i.e.} $\Delta\left.\partial_E\ln \nu(E)\right|_{\mu}\ll 1$. Using Eqs.~\eqref{eq:S_zvyagin_1}
and~\eqref{eq:p_zvyagin}, one finds 
\be\label{eq:S_zvyagin_2}
S\approx\frac{k_B}{e}\left(\frac{k_BT_M}{4}\right)\left.\partial_E\ln \nu(E)\right|_{\mu}\,.
\ee
This shows that the thermopower should be temperature independent when the assumptions above are fulfilled, 
which is always the case at very low temperatures (bottom part of region~(2a) in Fig.~\ref{Fig3}).  
Note that the same hypothesis for the DOS lead to the standard Mott formula~\eqref{eq:log_G_Mott} 
for the conductance: Eq.~\eqref{eq:S_zvyagin_2} describes the thermopower when Eq.~\eqref{eq:log_G_Mott} holds 
for the conductance.\\
\indent Let us now consider the case where the impurity band edges are explored, say the lower one.  
In analogy to the previous section, using a rough step-like model for $\nu(E)$ provides useful insight.  
Using Eq.~\eqref{eq:dos_theta} for the DOS and Eq. \eqref{eq:S_zvyagin_1}, one gets for the thermopower
\begin{subequations}
\begin{align}
S =\,\,& \frac{k_B}{e}\left(\frac{\epsilon_c-\mu}{2k_BT}+\frac{\Delta(T)}{2k_BT}\right)~~\mathrm{if}~~\epsilon_c<\mu~~\mathrm{and}~~\mu-\epsilon_c<\Delta, 
\label{eq:S_zvyagin_31}\\
S =\,\,& \frac{k_B}{e}\left(\frac{\epsilon_c-\mu}{k_BT}+\frac{\Delta(T)}{2k_BT}\right)~~\mathrm{if}~~\epsilon_c>\mu\,, \label{eq:S_zvyagin_32}
\end{align}
\end{subequations}
assuming\footnote{
We have also calculated the thermopower beyond this approximation, by plugging Eq.~\eqref{eq:dos_theta} 
for $\nu(E)$ into Eq.~\eqref{eq:p_zvyagin} for $p(E)$. Instead of Eqs.~\eqref{eq:S_zvyagin_31} 
and~\eqref{eq:S_zvyagin_32}, we find respectively
\begin{subequations}
\begin{align}
S =\,\,& \frac{k_B}{e}\left[\frac{5(\epsilon_c-\mu)}{8k_BT}+\frac{3\Delta(T)}{8k_BT}+O\left(\frac{\epsilon_c-\mu}{k_BT}\right)\right],\\
S =\,\,& \frac{k_B}{e}\left[\frac{\epsilon_c-\mu}{k_BT}+\frac{3\Delta(T)}{8k_BT}\right]~.
\end{align}
\end{subequations}
The two sets of equations are obviously very similar.  At a qualitative level of analysis, 
it is meaningless to favour one over the other.} $p(E)=1$ in the energy window $|E-\mu|<\Delta$ 
[$0<E-\epsilon_c<\Delta$] and $0$ elsewhere. Similar formulas can be deduced by symmetry 
if the upper band edge is explored.  The resulting thermopower behaviour as a function of 
temperature turns out to be rich.  Indeed, depending on the position of $\mu$ with respect to the 
(bottom) edge $\epsilon_c$ of the DOS, and depending on the magnitude of $\Delta$, 
the thermopower can be an increasing or decreasing function of $T$.  If $\mu$ lies outside the impurity band, 
the thermopower (in unit of $k_B/e$ if not otherwise specified) is found to be a monotonically 
decreasing function of the temperature (see Eq.~\eqref{eq:S_zvyagin_32}).  
On the other hand, if $\mu$ lies inside the band, 
close to the edge $\epsilon_c$ of the DOS, the thermopower increases with the temperature, reaches a maximum 
(at $k_BT=(\epsilon_c-\mu)^2/(16k_BT_M)$) and then starts to decrease (see Eq.~\eqref{eq:S_zvyagin_31}).\\
\indent Let us finally address the large temperature limit ($k_BT\gtrsim 2E_B$), corresponding to region~(4b) 
and the upper part of region~(4a) in Fig.~\ref{Fig3}.  In that case, all impurity 
band states are involved in thermoelectric transport, with $p(E)\approx 1$.  As a consequence, 
the thermopower temperature behaviour is merely $S\sim T^{-1}$.  Assuming a constant DOS, one gets
\be
\label{eq:S_largeT}
S=\frac{k_B}{e}\left(\frac{V_g-\mu}{k_BT}\right)\,.
\ee

\subsection{Numerical results}
\begin{figure}
  \centering
  \includegraphics[keepaspectratio, width=\columnwidth]{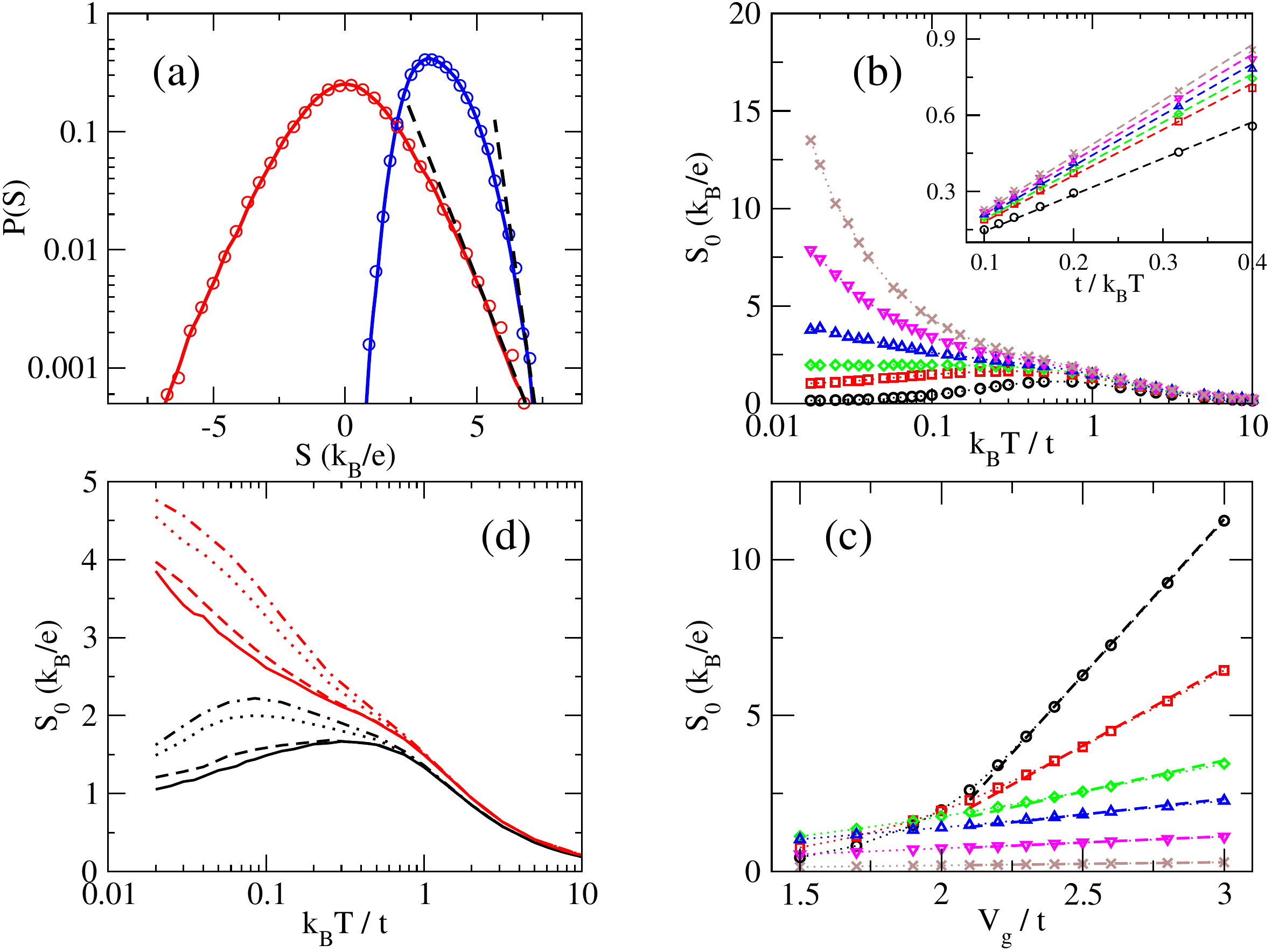}
  \caption{\label{Fig6} 
In all panels, unless specified, $L=200$, $\mu=0$, $W=t$ and $\gamma_e=\gamma_{ep}=t$. (a) Thermopower distributions in the VRH regime, 
when $\mu$ lies in the bulk (left red curve, $V_g=0$, $W=4t$) or close to the edge (right blue curve, $V_g=2.2t$, $W=t$) of the impurity 
band. Data are given for $L=200$ (full lines) and $L=400$ (circles). The straight dashed lines underline the exponential behaviour of 
the tails $\sim \text{exp}\{-cS\}$ predicted in Ref.~\cite{Jiang2013}. In both cases, $k_BT=t$. (b) Main panel: Typical thermopower as 
a function of $T$ around the (lower) band edge. From the bottom to the top, the various curves 
correspond to $V_g/t=1.5\,\text{({\large{$\circ$}})},1.9\,\text{({\scriptsize{\color{red}$\square$}})},
2.0\,\text{({\large{\color{ForestGreen}$\diamond$}})},2.1\,\text{({\scriptsize{\color{blue}$\triangle$}})},
2.2\,\text{({\scriptsize{\color{magenta}$\triangledown$}})}$  and $2.3\,\text{({\scriptsize{\color{brown}$\times$}})}$. 
Dotted lines are guides to the eye. Inset: zoom at very large temperatures $k_BT\gtrsim E_B$. The fits $f(V_g)/T$ (dashed lines) 
confirm the expected behaviour $S_0\sim T^{-1}$ (Eq.~\eqref{eq:S_largeT}). (c) Typical thermopower as a function of $V_g$, 
for $k_BT/t=0.1\,\text{({\large{$\circ$}})},0.2\,\text{({\scriptsize{\color{red}$\square$}})},
0.5\,\text{({\large{\color{ForestGreen}$\diamond$}})},1.0\,\text{({\scriptsize{\color{blue}$\triangle$}})},
2.5\,\text{({\scriptsize{\color{magenta}$\triangledown$}})}$ and $10.0\,\text{({\scriptsize{\color{brown}$\times$}})}$. 
At large $V_g$ (when $\mu$ lies outside the band), dashed lines are linear fits with slope $t/k_BT$ (Eq.~\eqref{eq:S_zvyagin_32}). 
Dotted lines are guides to the eye. (d) Typical thermopower as a function of $T$, for electron-phonon coupling strength $\gamma_{ep}/t=1$ 
(full line), $0.5$ (dashed line), $0.1$ (dotted line) and $0.05$ (mixed line), at $V_g=1.9t$ (black curves, bottom set) and $V_g=2.1t$ (red curves, 
top set).
}
\end{figure}
\indent For the model introduced in Sec.~\ref{sec:Andmodel},  we now study the thermopower by solving numerically the random resistor network 
(see \ref{app_resnet}).\\ 
\indent Fig.~\ref{Fig6}(a) gives the distribution $P(S)$ of the thermopower $S$ in the VRH regime, when the impurity band center (red curve) 
and lower edge (blue curve) are probed at $\mu$. While the thermopower distribution is symmetric around a vanishing average value at the band 
center, it is shifted away from $0$ and gets skewed close to the band edges.  Such features can be easily understood: 
The level distribution becomes highly asymmetric with respect to $\mu$ when one probes the lower band edge with a positive gate 
voltage $V_g$. Consequently, an electron entering the nanowire from the left lead around $\mu$ finds more states \textit{above} its 
energy than below. It has therefore a tendency to absorb energy in order to move to regions of higher DOS, before releasing it at 
the right side of the nanowire, as illustrated in Fig.~\ref{Fig2}. Recalling that $S=\langle E-\mu \rangle/(eT)$, one can thus 
explain why $P(S)$ is shifted and skewed at finite $V_g$.  Let us notice that such a skewness cannot be seen in the low-temperature 
coherent regime~\cite{Bosisio2014}, where transport only involves electrons at energies very close to $\mu$; In that case, 
distributions are found to be shifted with $V_g$ but always symmetric.  Another important message of Fig.~\ref{Fig6}(a) is 
that for both values of $V_g$ the thermopower distribution turns out to be independent of the nanowire length $L$. This is consistent  
with the observation that the thermopower is governed by the edges of the nanowire in the hopping regime, as recently 
pointed out in Ref.~\cite{Jiang2013}.\\
\indent We then investigate the typical thermopower behaviour as a function of temperature and gate voltage, by extracting the 
median $S_0$ of the distribution $P(S)$ for different sets of parameters.  The temperature dependence of $S_0$ is shown in 
Fig.~\ref{Fig6}(b) for different values of the gate voltage, which have been chosen for scanning the vicinity 
of the lower band edge. The main observation is that our model predicts a huge enhancement of the thermopower around the band 
edges. Values larger that $10\, k_B/e$ are obtained by properly tuning the strength of the gate voltage in the VRH regime.
Other features of those curves are worth emphasizing:
\begin{enumerate}
\item $S_0$ is always positive in unit of $k_B/e$, hence negative in $\mathrm{V\,K}^{-1}$ (since $e<0$).  
This is expected since transport is due to electrons near the lower band edge, the sign of the thermopower reflecting 
the sign of the charge carriers.\footnote{The occurrence of negative $S_0$ is nevertheless possible not far from the 
lower band edge, as soon as $\Delta$ is sufficiently small and the DOS slope at $\mu$ becomes strongly negative. 
In our model, such a negative slope occurs close to the band edges, as shown in Fig.~\ref{Fig4}.}
\item At low temperatures the typical thermopower can either increase or decrease 
with the temperature depending on the gate voltage. Roughly speaking, it increases inside the band and decreases 
outside, in agreement with the theoretical predictions~\eqref{eq:S_zvyagin_31} and~\eqref{eq:S_zvyagin_32}, 
obtained assuming a step-like model for the DOS close to the band edge $\epsilon_c$.  
Moreover, the position of the crossover between the two behaviours is found around $V_g-\mu\approx 2t$, 
a value consistent with our previous estimation of the (lower) band edge position of the Anderson model at $\epsilon_c\approx V_g-2.2t$ (see Sec.~\ref{subsec_Tscale}).
\item At high temperature (typically larger than the bandwidth), the curves converge to a $T^{-1}$ behaviour, as shown 
in the inset of Fig.~\ref{Fig6}(b). The crude estimation~\eqref{eq:S_largeT} turns out 
to be satisfactory in this regime.
\item In the low temperature limit and in the case where $\mu$ lies 
inside the band, the typical thermopower $S_0$ is expected to saturate, according to Eq.~\eqref{eq:S_zvyagin_2}. 
Such a saturation is not observed in Fig.~\ref{Fig6}(b).  Two reasons can be invoked.  
The first one is that Eq.~\eqref{eq:S_zvyagin_2} was actually derived under the assumption of a constant 
localisation length $\xi_i\approx \xi(\mu)$ while the numerical results reported here were obtained going 
beyond this approximation, by taking into account the energy dependency of the different 
localisation lengths $\xi_i$ of sites $i$. In \ref{app_matrixelement}, we show that under 
the assumption $\xi_i\approx \xi(\mu)$, $S_0$ indeed saturates at low temperature.  The other possibility 
is simply that the saturation appears at lower temperatures, which are not reachable numerically because 
of round-off errors.
\item For high values of $V_g$, the typical thermopower seems to diverge as the temperature 
is lowered. It is obvious that the thermopower eventually decreases below a certain temperature, 
since all curves in Fig.~\ref{Fig6}(b) are known to drop down to zero in the zero-temperature 
limit (linearly with $T$ and with a positive slope)~\cite{Bosisio2014}.
\end{enumerate}

In Fig.~\ref{Fig6}(c), we show how the typical thermopower depends on the gate voltage, 
for different values of the temperature. Approaching the edge of the impurity band, we see that $S_0$ increases, 
the effect being more pronounced at low temperatures. Outside the band, the behaviour of $S_0$ 
with $V_g$ is perfectly well fitted by the formula $S_0=(k_B/e)[\frac{V_g}{k_BT}+f(T)]$, 
as illustrated by the straight lines in Fig.~\ref{Fig6}(c). This linear enhancement of $S_0$ 
with $V_g$, as well as its range of validity, is consistent with the prediction~\eqref{eq:S_zvyagin_32} 
and our initial estimation $\epsilon_c\approx V_g-2.2t$ for the position of the lower band edge.  
Note however that Eq.~\eqref{eq:S_zvyagin_32} does not capture the $y$-intercept 
$f(T)\approx 0.89-1.94/(k_BT)$ of the linear fits.  On the other hand 
the fact that $S_0$ keeps increasing even outside the impurity band, when the conductance drops 
exponentially, may seem in contrast with recent experimental observations~\cite{Brovman2013}.  
We think the explanation lies in the fact that, when the nanowire is almost completely depleted by 
$V_g$, the probability for an electron at $\mu$ to tunnel inside the band becomes extremely small, 
and so do the electrical and heat currents; consequently, they may be too hard to measure.  
Nonetheless their ratio, which gives the thermopower, remains formally well defined and finite.\\
\indent We conclude our analysis by discussing the order of magnitude of our numerical results.  
In panels (a), (b) and (c) of Fig.~\ref{Fig6}, 
data was obtained taking $\gamma_e=\gamma_{ep}=t$ as input parameters of the model.  
In panel~(d) we investigate how the typical thermopower depends on the choice of these parameters,  
finding that $S_0$ does not vary by more than $50\%$ when the ratio $\gamma_e/\gamma_{ep}$ 
is increased or decreased by an order of magnitude.  
Remarkably, at the lowest studied temperatures (in the VRH regime) and around the band edges, 
the typical thermopower is found to reach very large values of the order of 
$10\,(k_B/e) \sim 1\,\mathrm{mV\,K}^{-1}$. It is worthwhile to note that, despite the simplicity of 
the model, the order of magnitude of these results is comparable to recent measurements of 
thermopower in semiconducting nanowires~\cite{Roddaro2013,Moon2013,Curtin2013,Brovman2013}, showing 
strong thermoelectric conversion at the band edges.


\section{Discussion and conclusion}
\label{section_ccl}

We have studied thermoelectric transport in a disordered nanowire in the field effect transistor configuration,
focusing on intermediate to high temperatures.  More precisely, $T$ was high enough for inelastic processes
(phonon-assisted hopping between localised states) to be dominant, but still such that $k_BT<2E_B$, with $2E_B$
the spread in available nanowire states.  Transport in this regime is typically of variable range hopping 
type~\cite{Mott1969,Mott1979}. We have extended the Miller-Abrahams random resistor network model~\cite{Miller1960} 
to deal with band-edge transport, and performed accurate numerical analysis based on the 1D Anderson model.  
The thermopower shows remarkable gate- and temperature-dependent behaviour, whose features can be
understood within a suitable generalization of Zvyagin's analytical treatment of VRH transport~\cite{Zvyagin1973,Zvyagin1991}.
In particular we have shown them to be largely independent of fine system details such as electron-phonon interaction 
strength or the specific form of the DOS.  Our results are in line with numerous experimental 
observations~\cite{Roddaro2013,Moon2013,Curtin2013,Brovman2013}, confirming the great thermoelectric conversion 
potential of band-edge transport. Notice in particular that semi-quantitative agreement with observations was reached, 
though we stress that our treatment's strength lies in its general applicability rather than in its high precision
-- the latter being heavily dependent on fine details of each particular setup, such as materials involved, 
doping level/type, geometry and so on.

Let us now comment on certain limitations of our work.
First, interactions have been neglected, except for the requirement of single-occupation of any given
localised state~\cite{Ambegaokar1971}.  Whereas this is appropriate in some cases, it is by no means
a universally valid assumption.  Indeed, numerous delicate issues related to the role of interactions
in activated transport are discussed in~\cite{Shklovskii1984} and references therein.
Secondly, we have ignored phonon-drag effects, which is however a much safer bet.
It is well known that the latter can play a prominent role in standard band 
transport -- i.e. when electronic states are delocalised -- 
but are irrelevant when transport is due to hopping between localised states\cite{Zvyagin1973,Zvyagin1991}.
Finally, we used the Anderson model which is a single band model and hence neglected the possibility of temperature activated 
transport via other bands. This amounts to assuming that $k_BT<E_{act}$, where $E_{act}$ is the interband spacing.  
$E_{act}$ depends on the considered material, ranging from tens of Kelvin degrees for weakly doped crystalline 
materials, to hundreds of Kelvin degrees in amorphous materials~\cite{Shklovskii1984}.

\ack
This work has been supported by CEA through the DSM-Energy Program (project E112-7-Meso-Therm-DSM). We acknowledge 
stimulating discussions with Y. Imry, R. Jalabert, F. Ladieu, K. Muttalib and A. Parola.


\appendix
\section{Solution of the random resistor network} 
\label{app_resnet}
In linear response we assume that on each localised state the electron occupation is characterized 
by a \emph{local distribution}~\cite{Shklovskii1984,Jiang2013}:
\be
f_i=f^0_i+\delta f_i,
\ee
where $f_i^0$ is the Fermi distribution at equilibrium (i.e., evaluated at the reference values $\mu$ and $T$), and 
$\delta f_i$ is the correction induced by the (small) applied bias $\delta\mu$. Linearizing Eqs.\eqref{eq:currents}, 
and making use of Eqs.~\eqref{eq:gamma_ij},~\eqref{eq:gamma_ij_2}, and~\eqref{eq:gamma_il}, the hopping currents 
between each pair of localised states, and the tunnelling currents from/to the electrodes can be written in terms of 
``local potentials'' $U_i$'s:
\begin{align}\label{eq:currents_app}
    &I_{ij} = G_{ij} (U_i-U_j),\nonumber\\
    &I_{iL(R)} = G_{iL(R)} (U_i-U_{L(R)}),
\end{align}
where
\begin{align}\label{eq:parameters_app}
    &G_{ij} = \frac{e^2}{k_BT}\,\gamma_{ij} f_i^0(1-f_j^0)(N_{ij}+1/2\mp 1/2),\cr
    &G_{iL(R)} = \frac{e^2}{k_BT}\,\gamma_{iL(R)} f_i^0(1-f_i^0),\cr
    &U_{i} = \frac{k_BT}{e}\,\delta f_i/[f_i^0(1-f_i^0)],\cr
    &U_{L(R)}(E_i) = \frac{k_BT}{e}\,\delta f_{L(R)}/[f_i^0(1-f_i^0)].
\end{align}
In the above expressions, in case of double signs, the upper (lower) sign refers to $E_j>E_i$ ($E_j<E_i$).\\
At steady state, according to Kirchoff's conservation law, the net electric current throughout every node \emph{i} must vanish:
\begin{equation}\label{eq:kirchoff}
\left(\sum_{j\neq i} I_{ij}\right) + I_{iL} + I_{iR}=0.
\end{equation}
By plugging Eqs.~\eqref{eq:currents_app}, we end up with a set of $L$ equations (one for every node $i$) to calculate the $L$ local 
potentials $U_i$'s, which can be written conveniently in the matrix form:
\begin{equation}\label{eq:system}
\sum_j A_{ij} U_j = z_i,
\end{equation}
where
\begin{align}\label{eq:system2}
		A_{ij}&=-G_{ij}\qquad \text{(for $i\neq j$)},\cr
    A_{ii}&=\sum_{k\neq i}G_{ik} + G_{iL}+ G_{iR},\cr
    z_i&=G_{iL}\,\left(\delta\mu_L/e\right)
\end{align}
In writing the expression for $z_i$, we exploited the fact that $\delta \mu_R=\delta T_R=0$, having chosen to set the right terminal as 
reference (see Sec.~\ref{section_transport}).\\
Once the system is solved and the $U_i$ are known, all the $I_{ij}$'s and $I_{iL(R)}$  can be calculated via Eqs. \eqref{eq:currents_app}. 
The electric and heat current can be computed by summing the outgoing contributions from the left (right) lead toward \emph{every} states in the system:
\begin{align}\label{eq:currents_app2}
& I^e_L=-\sum_i I_{iL}=\sum_i I_{iR},\cr
& I^Q_{L(R)}= \sum_i\left(\frac{E_i-\mu_{L(R)}}{e}\right)I_{L(R)i}.
\end{align}

\section{Calculation of the hopping probability} 
\label{app_matrixelement}
Miller and Abrahams\cite{Miller1960,Shklovskii1984} described how to calculate the hopping probability $\gamma_{ij}$ between two donors $i$ and $j$ 
in a 3D semiconductor, mediated by the absorption or emission of a phonon. When the distance between the donors is large, they obtain for 
$\gamma_{ij}$ an expression which depends on the (weak) overlap between the donor wavefunctions and on the mutual electrostatic effect between them: 
\be
\gamma_{ij}\propto \left|\langle \psi_i| \frac{e^2}{\kappa |\boldsymbol{r}-\boldsymbol{r}_i|}|\psi_j\rangle - \langle \psi_i| 
\psi_j\rangle \langle \psi_i| \frac{e^2}{\kappa |\boldsymbol{r}-\boldsymbol{r}_j|}|\psi_i\rangle \right|^2.
\label{eq:gamma_ij_app}
\ee
If the donor wavefunctions $\psi_i$ and $\psi_j$ are characterized by the \emph{same} decay length $\xi$, Eq.~\eqref{eq:gamma_ij_app} can be 
simplified~\cite{Miller1960,Shklovskii1984} 
\be
\gamma_{ij} \propto \exp (-2|\boldsymbol{r}_i-\boldsymbol{r}_j|/\xi).  
\label{eq:gamma_ij_app_2}
\ee
If the decay lengths of $\psi_i$ and $\psi_j$ are different ($\xi_i\neq\xi_j$), a rigourous evaluation of $\gamma_{ij}$ from Eq.~\eqref{eq:gamma_ij_app} may be complicated, but the key point is that it will always be proportional to the overlap $\langle \psi_i|\psi_j\rangle$. Hence, we can write it in the form
\be
\gamma_{ij}\propto \left| \langle \psi_i| \psi_j\rangle \right|^2 \sim \left|\mathcal{C}_i \exp (-r_{ij}/\xi_i) 
+ \mathcal{C}_j \exp (-r_{ij}/\xi_j )\right|^2,
\label{eq:gamma_ij_app_3}
\ee
where $r_{ij}=|\boldsymbol{r}_i-\boldsymbol{r}_j|$ is the distance between $i$ and $j$, and the coefficients $\mathcal{C}_i$ and 
$\mathcal{C}_j$ depend on $\xi_i$, $\xi_j$ and $r_{ij}$. The explicit form of these coefficients will take into account all details 
concerning the wavefunction overlap $\langle \psi_i| \psi_j\rangle$. In 1D the calculation becomes simpler and leads to 
Eq.~\eqref{eq:gamma_ij_app_4}. Extending a theory originally developed for lightly doped cristalline semiconductors (where the decay length 
is the donor Bohr radius) to Anderson insulators (where the decay length becomes the localisation length), 
Ambegaokar \emph{et al.}\cite{Ambegaokar1971} have used Eq.~\eqref{eq:gamma_ij_app_2} for describing the hopping probability.
For similar reasons, we use Eq.~\eqref{eq:gamma_ij_app_4} in our numerical calculations, for both $G$ and $S$, taking for $\xi_i$ and $\xi_j$ 
the localisation length of two Anderson localised states.

\begin{figure}
  \centering
  \includegraphics[keepaspectratio, width=\columnwidth]{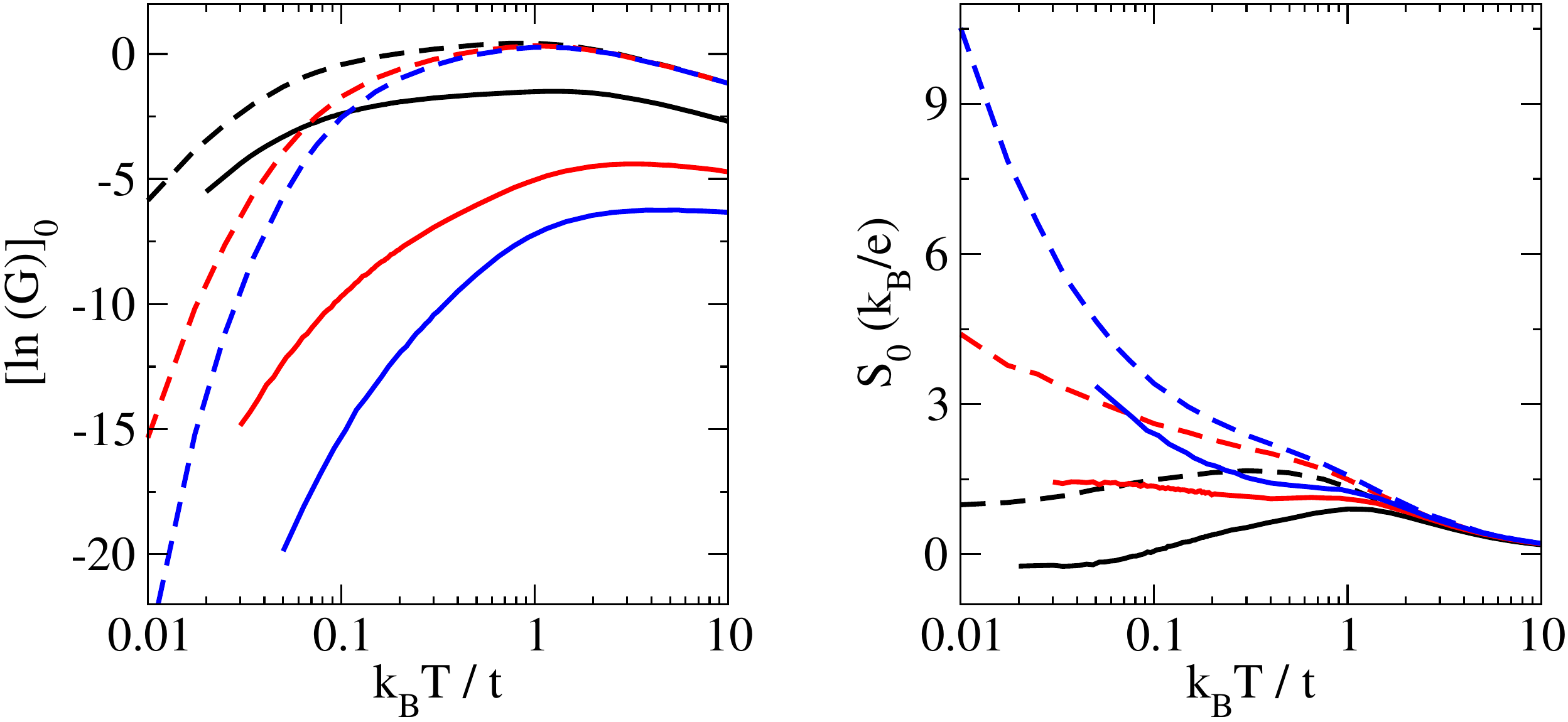}
  \caption{\label{Fig7} 
  Typical logarithm of electrical conductance (left) and typical thermopower (right) of a disordered nanowire as a function of temperature, 
and for different values of the applied gate voltage. Full lines refer to the approximation in which all $\xi_i=\xi(\mu)$, while dashed lines refer to 
the improved theory in which the energy dependence of $\xi_i=\xi(E_i)$ is taken into account for evaluating the transition rates. In each set, from 
the top to the bottom (left panel) or reversely (right panel), the curves correspond to $V_g=1.9t$ (black), $V_g=2.1t$ (red) and $V_g=2.2t$ (blue). 
Other parameters: $L=200$, $W=t$, $\mu=0$, $\gamma_e=\gamma_{ep}=t$.}
\end{figure}

\indent In order to estimate the difference between taking $\xi(\mu)$ or $\xi(E)$ when computing the transition rates (Eqs.\eqref{eq:gamma_ij} 
and \eqref{eq:gamma_il}), we have calculated the typical logarithm of the conductance and the typical thermopower as functions of the 
temperature in the two cases: Fig.~\ref{Fig7} shows that there is no \emph{qualitative} difference between the curves 
computed using $\xi(\mu)$ (full lines) and $\xi(E)$ (dashed lines). The main effect of taking into account the localisation length energy 
dependence is that, according to Eq.\eqref{eq:gamma_ij_app_4}, all transitions toward the more delocalised states around the band center are 
favoured. This leads to a much better conductance especially at low temperatures, where the difference could be of several orders of magnitude; 
on the other hand, the effect on the thermopower is weaker.


\section*{References}
\bibliographystyle{iopart-num}
\bibliography{thermo1dVRH}

\providecommand{\newblock}{}
\begin{thebibliography}{10}
\expandafter\ifx\csname url\endcsname\relax
  \def\url#1{{\tt #1}}\fi
\expandafter\ifx\csname urlprefix\endcsname\relax\def\urlprefix{URL }\fi
\providecommand{\eprint}[2][]{\url{#2}}

\bibitem{Chen2003}
Chen G, Dresselhaus M~S, Dresselhaus G, Fleurial J~P and Caillat T 2003 {\em
  Int. Mater. Rev.\/} {\bf 48} 45

\bibitem{Hicks1993}
Hicks L~D and Dresselhaus M~S 1993 {\em Phys. Rev. B\/} {\bf 47} 16631

\bibitem{Curtin2012}
Curtin B~M, Fang E~W and Bowers J~E 2012 {\em J. Electron. Mat.\/} {\bf 41} 887

\bibitem{Blanc2013}
Blanc C, Rajapbour A, Volz S, Fournier T and Bourgeois O 2013 {\em Appl. Phys.
  Lett.\/} {\bf 103} 043109

\bibitem{Brovman2013}
Brovman Y~M, Small J~P, Hu Y, Fang Y, Lieber C~M and Kim P 2013 {\em
  arXiv:1307.0249\/}

\bibitem{Stranz2013}
Stranz A, Waag A and Peiner E 2013 {\em J. Electron. Mat.\/} {\bf 42} 2233

\bibitem{Karg2013}
Karg S, Mensch P, Gotsmann B, Schmid H, Kanungo P~D, Ghoneim H, Schmidt V,
  Bj\"{o}rk M~T, Troncale V and Riel H 2013 {\em J. Electron. Mat.\/} {\bf 42}
  2409

\bibitem{Roddaro2013}
Roddaro S, Ercolani D, Safeen M~A, Suomalainen S, Rossella F, Giazotto F, Sorba
  L and Beltram F 2013 {\em Nano Lett.\/} {\bf 13} 3638

\bibitem{Hochbaum2008}
Hochbaum A~I, Chen R, Delgado R~D, Liang W, Garnett E~C, Najarian M, Majumdar A
  and Yang P 2008 {\em Nature\/} {\bf 451} 163

\bibitem{Tilke2002}
Tilke A, Pescini L, Erbe A, Lorenz H and Blick R~H 2002 {\em Nanotechnology\/}
  {\bf 13} 491

\bibitem{Bourgeois2007}
Bourgeois O, Fournier T and Chaussy J 2007 {\em J. Appl. Phys.\/} {\bf 101}
  016104

\bibitem{Boukai2008}
Boukai A~I, Bunimovich Y, Tahir-Kheli J, Yu J~K, Goddard W~A and Heath J~R 2008
  {\em Nature\/} {\bf 451} 168

\bibitem{Galli2010}
Galli G and Donadio D 2010 {\em Nat. Nanotech.\/} {\bf 5} 701

\bibitem{Heron2010}
Heron J~S, Bera C, Fournier T, Mingo N and Bourgeois O 2010 {\em Phys. Rev.
  B\/} {\bf 82} 155458

\bibitem{He2012}
He Y and Galli G 2012 {\em Phys. Rev. Lett.\/} {\bf 108} 215901

\bibitem{Hu2012}
Hu M and Poulikakos D 2012 {\em Nano Lett.\/} {\bf 12} 5487

\bibitem{Jiang2012}
Jiang J~H, Entin-Wohlman O and Imry Y 2012 {\em Phys. Rev. B\/} {\bf 85} 075412

\bibitem{Jiang2013}
Jiang J~H, Entin-Wohlman O and Imry Y 2013 {\em Phys. Rev. B\/} {\bf 87} 205420

\bibitem{Bosisio2014}
Bosisio R, Fleury G and Pichard J~L 2014 {\em New J. Phys.\/} {\bf 16} 035004

\bibitem{Zvyagin1973}
Zvyagin I~P 1973 {\em Phys. Stat. Sol. (b)\/} {\bf 58} 443

\bibitem{Zvyagin1991}
Zvyagin I~P 1991 {\em Hopping Transport in Solids\/} (ed. by M. Pollak and B.
  I. Shklovskii (North-Holland, Amsterdam))

\bibitem{Poirier1999}
Poirier W, Mailly D and Sanquer M 1993 {\em Phys. Rev. B\/} {\bf 59} 10856

\bibitem{Tilke1999}
Tilke A, Blick R~H, Lorenz H, Kotthaus J~P and Wharam D~A 1999 {\em Appl. Phys.
  Lett.\/} {\bf 75} 3704

\bibitem{Dayen2009}
Dayen J~F, Wader T~L, Rizza G, Golubev D~S, Cojocaru C~S, Pribat D, Jehl X,
  Sanquer M and Wegrowe J~E 2009 {\em Eur. Phys. J. Appl. Phys.\/} {\bf 48}
  10604

\bibitem{Rodin2010}
Rodin A~S and Fogler M~M 2010 {\em Phys. Rev. Lett.\/} {\bf 105} 106801

\bibitem{Ambegaokar1971}
Ambegaokar V, Halperin B~I and Langer J~S 1971 {\em Phys. Rev. B\/} {\bf 4}
  2612

\bibitem{Miller1960}
Miller A and Abrahams E 1960 {\em Phys. Rev.\/} {\bf 120} 745

\bibitem{Mott1969}
Mott N~F 1969 {\em Phil. Mag.\/} {\bf 19} 835

\bibitem{Mott1979}
Mott N~F and Davis E~A 1979 {\em Electronic Processes in Non Crystalline
  Materials\/} (Clarendon, Oxford, (2nd ed.))

\bibitem{Kurkijarvi1973}
Kurkij\"arvi J 1973 {\em Phys. Rev. B\/} {\bf 8} 922

\bibitem{Raikh1989}
Raikh M~E and Ruzin I~M 1989 {\em Sov. Phys. JETP\/} {\bf 68} 642

\bibitem{Serota1986}
Serota R~A, Kalia R~K and Lee P~A 1986 {\em Phys. Rev. B\/} {\bf 33} 8441

\bibitem{Pollack1972}
Pollack M 1972 {\em J. Non-Cryst. Solids\/} {\bf 11} 1

\bibitem{Shklovskii1984}
Shklovskii B and Efros A 1984 {\em Electronic Properties of Doped
  Semiconductors\/} (Springer-Verlag, Berlin)

\bibitem{Callen1985}
Callen H 1985 {\em Thermodynamics and an Introduction to Thermostatics\/} (John
  Wiley and Sons, New York)

\bibitem{Derrida1984}
Derrida B and Gardner E 1984 {\em J. Physique\/} {\bf 45} 1283

\bibitem{Lee1984}
Lee P~A 1984 {\em Phys. Rev. Lett.\/} {\bf 53} 2042

\bibitem{Kwasnick1984}
Kwasnick R~F, Kastner M~A, Melngailis J and Lee P~A 1984 {\em Phys. Rev.
  Lett.\/} {\bf 52} 224

\bibitem{Webb1985}
Webb R~A, Hartstein A, Wainer J~J and Fowler A~B 1985 {\em Phys. Rev. Lett.\/}
  {\bf 54} 1577

\bibitem{Ladieu1993exp}
Ladieu F, Mailly D and Sanquer M 1993 {\em J. Phys. I France\/} {\bf 3} 2321

\bibitem{Hasko1993}
Hasko D~G, Cleaver J~R~A, Ahmed H, Smith C~G and Dixon J~E 1993 {\em Appl.
  Phys. Lett.\/} {\bf 62} 2533

\bibitem{Han2010}
Han M~Y, Brant J~C and Kim P 2010 {\em Phys. Rev. Lett.\/} {\bf 104} 056801

\bibitem{Ladieu1993th}
Ladieu F and Bouchaud J~P 1993 {\em J. Phys. I France\/} {\bf 3} 2311

\bibitem{Rodin2011}
Rodin A~S and Fogler M~M 2011 {\em Phys. Rev. B\/} {\bf 84} 125447

\bibitem{Grant1974}
Grant A~J and Davis E~A 1974 {\em Sol. State Comm.\/} {\bf 15} 563

\bibitem{Rodin2009}
Rodin A~S and Fogler M~M 2009 {\em Phys. Rev. B\/} {\bf 80} 155435

\bibitem{Moon2013}
Moon J, Kim J~H, Chen Z, Xiang J and Chen R 2013 {\em Nano Lett.\/} {\bf 13}
  1196

\bibitem{Curtin2013}
Curtin B~M, Codecido E~A, Kr\"amer S and Bowers J~E 2013 {\em Nano Lett.\/}
  {\bf 13} 5503

\end{thebibliography}

\end{document}